\documentstyle[epsf]{elsart}
\newcommand{\postscript}[2] {\setlength{\epsfxsize}{#2\hsize}
\centerline{\epsfbox{#1}}}
\begin{document}
\vspace*{-1cm}
\normalsize
\hspace*{6.5cm}\parbox{15cm}{Accepted in Nucl.Phys.{\bf A} (1997)}
\vspace*{0.1cm}
\begin{frontmatter}
\title{Quark-antiquark correlation in the pion}
\author[USP]{J.P.B.C. de Melo}, 
\author[ITA]{T. Frederico}, 
\author[IFT]{Lauro Tomio} and
\author[IFT,JINR]{A. E. Dorokhov}
\address[USP]{Instituto de F\'\i sica, Universidade de S\~ao Paulo,
01498-970, S\~ao Paulo, Brazil} 
\address[ITA]{Departamento de F\'\i sica, ITA, Centro T\'ecnico
Aeroespacial, 12.228-900, S\~ao Jos\'e dos Campos, Brazil}
\address[IFT]{Instituto de F\'\i sica Te\'orica, UNESP,
Rua Pamplona, 145, 01405-900, S\~ao Paulo, Brazil}
\address[JINR]{Bogoliubov Theoretical Laboratory, Joint Institute for
Nuclear Research, 141980, Dubna, Russia}
\date{\today}
\begin{abstract}
The electromagnetic tensor for inclusive electron scattering off the
pion, ($W^{\mu\nu}$), for momentum transfers such that $q^+=0$,
($q^+=q^0+q^3$) is shown to obey a sum-rule for the component $W^{++}$.
>From this sum-rule, one can define the quark-antiquark correlation
function in the pion, which characterizes the transverse distance
distribution between  the quark and antiquark in the light-front pion
wave-function. Within the realistic models of the relativistic pion
wave function (including instanton vacuum inspired wave function) it is
shown that the value of the two-quark correlation radius ($r_{q\bar
q}$) is near twice the pion electromagnetic radius ($r_{\pi}$), where
$r_{\pi} \approx 2/3$ fm.  We also define the correlation length
$l_{corr}$ where the two - particle correlation have an extremum. The
estimation of $l_{corr}\approx 0.3 - 0.5$ fm is very close to
estimations from instanton models of QCD vacuum.  It is also shown that
the above correlation is very sensitive to the pion light-front
wave-function models.
\end{abstract}
\begin{keyword}
Electromagnetic form factors,
relativistic quark model,
sum rules, pion
\end{keyword}
\end{frontmatter}
\vskip 0.2cm

Investigation of low-energy pion constants and pion form-factors at low
and intermediate momentum transfers provides important information
about internal dynamics of hadron constituents. \ At asymptotically
high momentum transfers the behaviour of pion form-factors is defined
by quark counting rules
\cite{brods1,mmt} and perturbative QCD gives rigorous predictions for
exclusive amplitudes \cite{EfRad,BrLep}.  However, some time ago the
applicability of the perturbative approach to exclusive processes at
moderately high momentum transfers has been stood under question
\cite{ILS,RadNPA}. It turns out that the attempts to describe the pion
form factor using only perturbative hard scattering mechanism is not
successful and soft internal dynamics of pion constituents becomes
important.  Later on, in refs.(\cite{li,jakro,doro}), it was shown that
it is necessary to include an intrinsic transverse momentum dependence
of the soft pion wave function to justify perturbative QCD calculations
of the pion form factors in the region of momentum transfers far below
of asymptotic one.  Moreover, numerical analysis show that at low and
intermediate momentum transfers, $Q < 3 - 5$ GeV, the soft
(overlapping) diagram dominates over the asymptotic (one - gluon
exchange) ones, inspite of the fact that at large $Q$ the first one is
parametrically smaller by $1/Q^2$.

So, detailed theoretical input and additional experimental information
is needed to relate low  and high energy properties of pion.  In this
work we want to consider a correlation function describing quark -
antiquark correlations in transverse space direction that could be in
principle measured in the electron inclusive scattering off pion at
moderately high energy experiments.  Considering the actual interest
and convenience in applying the light cone formalism to investigate the
hadronic structure at low and intermediate energies, it will be
important to derive some useful sum rules in this new context. \ We
derive for the pion a sum rule for the light cone component of the
inclusive hadronic tensor $W^{++}$ that is diagonal in the Fock state
basis, which gives the quark - antiquark correlation.  Also, applying
this sum rule, we study a few relativistic models for the pion wave-function.

Our basic assumption is that, at energy scale less then few GeV, 
exists a simple constituent quark
wave-function containing all the relevant physical information.
However, the relation between size, excitation spectrum and quark 
correlation in the hadron may be very complex.
We suppose that the pion is a strongly bound system of constituent 
quarks of masses 250 - 350 MeV.
In this picture, hadron amplitudes describe the transition of hadron
states into quark - antiquark pairs.  They are of nonperturbative
origin and  serve as absolute normalization (initial condition) of the
large $Q^2$ behaviour calculated perturbatively.

We will use the light
- cone constituent quark models with wave-functions defined in the
null-plane hypersurface ($x^+=x^0+x^3=0$)\cite{teren}.  This approach 
allows a consistent truncation of the Fock-space, 
such that the boost transformations that keep the
null-plane invariant, do not mix different Fock-components \cite{Weinb}
(see \cite{wil90} for a modern discussion of this problem). 
\ In that respect, we can work with a fixed number of constituents quarks.

Using constituent quarks degrees of freedom, 
the normalization of the wave function is well defined. It is finite. 
This fact does not forbid that the number of partons grows to infinite. In the 
limit of $x\to 0$,  the constituent $F_2(x)$, without considering 
the  constituent quark structure,  goes to zero. 
However, a description, with  constituents-with-structure, has 
been shown \cite{alta} to provide a reasonably accurate description of 
the experimental    deep-inelastic structure 
function of the pion. The partonic structure function of the 
constituent quark is convoluted with the  structure function obtained 
from the constituent quark wave-function. In this reference\cite{alta}, 
the experimental observation of $F_2(x=0)$  
being non-zero  is due to the constituent quark structure 
in terms of the partons. 

All light-cone operators, corresponding to physical quantities, are
classified as ``good'' and ``bad'' ones, where the ``good'' operators
are diagonal in the Fock state basis, as a consequence of suppression
of pair creation processes~\cite{Weinb,dash}. \ We show that, by using
the ``good - good'' component of the inclusive hadronic tensor,
$W^{++}$, for momentum transfers such that $q^+= 0$ and integrating in
$q^-$, it is possible to introduce a sum rule for $W^{++}$, which is
equal to the well known sum-rule presented in ref.~\cite{west}. 
The sum rule should approach the deep inelastic sum-rule at few GeV,
which diverges in the limit of $q^2\to \infty$.
For this reason we will consider the difference between 
the sum rules for charged and non-charged pions, where the
divergent part is cancelled, which permits to the define  
the constituent quark - antiquark density in the pion. 

The sum rule is the relativistic generalization 
of the Coulomb sum rule for inelastic
electron scattering. \ It is well known~\cite{Gott,wale} that the
Coulomb sum rule integral is the  Fourier transform of the two-body
density.  As its non-relativistic counterpart, we show that the
relativistic sum rule defines the correlation function characterizing
the transverse distance distribution of quark and antiquark in the
pion.  This allows a simple interpretation of the observable in terms
of constituent $q\overline q$ composite light-front pion
wave-function.

The electromagnetic tensor, $W^{\mu\nu}$, for the inclusive electron
scattering off pion, is defined as the square of the amplitude for the
photon absorption summed over all final hadron states:
\begin{eqnarray}\noindent
W^{\mu\nu}(p,q) &=& \frac{(2\pi)^4}{m_\pi} \sum_{n} \int
\frac{d^4x}{2\pi} \prod_{i=1}^{n} \left(\frac{d^3p_i}{(2\pi)^3 p_{i0}}
\right) e^{iqx}\times  \nonumber\\
&&\langle p_\pi | J^\gamma_\mu (x) |n\rangle  \langle n|
J^\gamma_\nu (0) |p_\pi\rangle \delta^4(p_n-p_\pi -q)\nonumber \\ 
&=&\frac{1}{m_\pi} \int \frac{d^4x}{2\pi} e^{iqx} \langle p_\pi |
J^\gamma_\mu (x) J^\gamma_\nu (0) |p_\pi\rangle ,
\label{wgam}
\end{eqnarray}
where $p$ denotes the four vector of the pion and $q$ is the photon
momentum transfer.  As a tensor of second rank  it is written in terms
of two invariant structure functions $W_1$ and $W_2$, as follows from
Lorentz, gauge and parity symmetries:
{\small
\begin{equation}\noindent
W^{\mu\nu}(p,q) = W_1(q^2,q.p)\left[\frac{q^\mu
q^\nu}{q^2}-g^{\mu\nu}\right] +
\frac{W_2(q^2,q.p)}{m_\pi^2}
\left[p^\mu - \frac{{ p. q} q^\mu}{q^2}\right]
\left[p^\nu - \frac{{ p. q} q^\nu}{q^2}\right].
\label{w}
\end{equation} }
In the rest frame of the pion, $p^+=p^-=p^0 = m_\pi$
and ${\bf p} = 0$, the photon momentum can be chosen such that
$q^+=0$, and the component $W^{++}$ is given by:

\begin{eqnarray}
W^{++}(p,q) \ =  \ W_2(q^2,q.p) \ .
\label{w+}
\end{eqnarray}

We shall consider the sum rule $C(q^2_\perp)$, by integrating $W^{++}$
in $q^-$ at fixed $q_\bot^2$ and $q^+=0$\footnote{It is assumed that
permutation between $q^+=0$ and $\int\limits_{0}^{\infty}dq^-$ is
allowed.}:

\begin{eqnarray}
C(q^2_\perp) = 
\frac{1}{2 e^2 }\int_{0}^{+\infty} dq^- W^{++}(p,q)|_{(q^+ = 0)} =
\frac{1}{2 e^2 }\int_{0}^{+\infty} dq^- W_2(q^2, q.p)|_{(q^+ = 0)} ,
\label{c}
\end{eqnarray}
where $e$ is the electron charge. In the kinematics we have chosen,
$q_\perp^2 = - q^2$ and $q^-/2 = \nu = p.q/m_\pi$, thus 
Eq.(\ref{c}) is precisely the sum-rule $ \int d\nu W_2( q^2,\nu)$
 at fixed $q^2$, defined in ref. \cite{west}. 
A similar sum-rule was first introduced by Gottfried \cite{Gott},
where $W^{00}$ is integrated over the transferred energy,
which is suitable for the use with instant form  wave-functions.

The well known Dashen - Gell - Mann - Fubini current algebra sum rule
(see for example, the book of De Alfaro et al. in ref.~\cite{dash})
differs from that suggested in Eq.(\ref{c}) in that respect that the
first one deal with the integration of the current commutator over
$q^-$ in the interval from $-\infty$ to $+\infty$. \ Due to crossing
symmetry, it becomes trivial in the case of electron inelastic
scattering and provides very important restrictions in the case of
neutrino scattering. In the case of the sum rule given in Eq.(\ref{c}),
the integration over $q^-$ is performed in the half axis interval,
$q^->0$, where $W^{\mu\nu}(p,q)$ is not equal to zero, so it is not
dominated by light - cone current algebra contribution.

It is well known that this kind of sum rule is unusual since it is of
``wrong" signature. Really, the derivation of the sum rules of ``right"
signature is based on consideration of causal amplitude which is
defined via time - ordered product of currents.  In that case the
absorptive part of the amplitude is expressed through the commutator of
currents satisfying causality principle. However, the sum rules of
``wrong" signature such as the Gottfried sum rule are constructed from
amplitudes of opposite crossing symmetry properties and correspond to
the matrix elements of the anti - commutators of currents \cite{Fox}.
Singular on the light cone, the contributions of the current anti -
commutator  provide parton sum rules (possessing scaling at $q^2 \to
\infty$) and describe the $SU(3)$ structure of hadrons. We shall
consider not only  singular contributions but also regular two -
particle contributions on the light cone, which describe correlations
in transverse space and correspond to power corrections to parton sum
rules.
\begin{figure}
\unitlength 0.8mm
\linethickness{1.0pt}
\begin{picture}(157.33,119.66)
\multiput(25.00,95.66)(-0.23,0.11){8}{\line(-1,0){0.23}}
\multiput(23.13,96.57)(-0.11,0.13){7}{\line(0,1){0.13}}
\multiput(22.36,97.48)(0.11,0.30){3}{\line(0,1){0.30}}
\multiput(22.69,98.39)(0.21,0.12){11}{\line(1,0){0.21}}
\multiput(25.00,99.66)(0.23,0.11){8}{\line(1,0){0.23}}
\multiput(26.87,100.57)(0.11,0.13){7}{\line(0,1){0.13}}
\multiput(27.64,101.48)(-0.11,0.30){3}{\line(0,1){0.30}}
\multiput(27.31,102.39)(-0.21,0.12){11}{\line(-1,0){0.21}}
\multiput(25.00,103.66)(-0.23,0.11){8}{\line(-1,0){0.23}}
\multiput(23.13,104.57)(-0.11,0.13){7}{\line(0,1){0.13}}
\multiput(22.36,105.48)(0.11,0.30){3}{\line(0,1){0.30}}
\multiput(22.69,106.39)(0.21,0.12){11}{\line(1,0){0.21}}
\multiput(25.00,107.66)(0.23,0.11){8}{\line(1,0){0.23}}
\multiput(26.87,108.57)(0.11,0.13){7}{\line(0,1){0.13}}
\multiput(27.64,109.48)(-0.11,0.30){3}{\line(0,1){0.30}}
\multiput(27.31,110.39)(-0.21,0.12){11}{\line(-1,0){0.21}}
\multiput(25.00,111.66)(-0.23,0.11){8}{\line(-1,0){0.23}}
\multiput(23.13,112.57)(-0.11,0.13){7}{\line(0,1){0.13}}
\multiput(22.36,113.48)(0.11,0.30){3}{\line(0,1){0.30}}
\multiput(22.69,114.39)(0.21,0.12){11}{\line(1,0){0.21}}
\multiput(25.00,115.66)(0.23,0.11){8}{\line(1,0){0.23}}
\multiput(26.87,116.57)(0.11,0.13){7}{\line(0,1){0.13}}
\multiput(27.64,117.48)(-0.11,0.30){3}{\line(0,1){0.30}}
\multiput(27.31,118.39)(-0.21,0.12){11}{\line(-1,0){0.21}}
\multiput(55.00,95.33)(-0.23,0.11){8}{\line(-1,0){0.23}}
\multiput(53.13,96.24)(-0.11,0.13){7}{\line(0,1){0.13}}
\multiput(52.36,97.15)(0.11,0.30){3}{\line(0,1){0.30}}
\multiput(52.69,98.06)(0.21,0.12){11}{\line(1,0){0.21}}
\multiput(55.00,99.33)(0.23,0.11){8}{\line(1,0){0.23}}
\multiput(56.87,100.24)(0.11,0.13){7}{\line(0,1){0.13}}
\multiput(57.64,101.15)(-0.11,0.30){3}{\line(0,1){0.30}}
\multiput(57.31,102.06)(-0.21,0.12){11}{\line(-1,0){0.21}}
\multiput(55.00,103.33)(-0.23,0.11){8}{\line(-1,0){0.23}}
\multiput(53.13,104.24)(-0.11,0.13){7}{\line(0,1){0.13}}
\multiput(52.36,105.15)(0.11,0.30){3}{\line(0,1){0.30}}
\multiput(52.69,106.06)(0.21,0.12){11}{\line(1,0){0.21}}
\multiput(55.00,107.33)(0.23,0.11){8}{\line(1,0){0.23}}
\multiput(56.87,108.24)(0.11,0.13){7}{\line(0,1){0.13}}
\multiput(57.64,109.15)(-0.11,0.30){3}{\line(0,1){0.30}}
\multiput(57.31,110.06)(-0.21,0.12){11}{\line(-1,0){0.21}}
\multiput(55.00,111.33)(-0.23,0.11){8}{\line(-1,0){0.23}}
\multiput(53.13,112.24)(-0.11,0.13){7}{\line(0,1){0.13}}
\multiput(52.36,113.15)(0.11,0.30){3}{\line(0,1){0.30}}
\multiput(52.69,114.06)(0.21,0.12){11}{\line(1,0){0.21}}
\multiput(55.00,115.33)(0.23,0.11){8}{\line(1,0){0.23}}
\multiput(56.87,116.24)(0.11,0.13){7}{\line(0,1){0.13}}
\multiput(57.64,117.15)(-0.11,0.30){3}{\line(0,1){0.30}}
\multiput(57.31,118.06)(-0.21,0.12){11}{\line(-1,0){0.21}}
\multiput(15.00,85.66)(0.12,0.13){17}{\line(0,1){0.13}}
\multiput(17.02,87.94)(0.12,0.12){17}{\line(0,1){0.12}}
\multiput(19.04,90.02)(0.13,0.12){16}{\line(1,0){0.13}}
\multiput(21.06,91.92)(0.13,0.11){15}{\line(1,0){0.13}}
\multiput(23.08,93.62)(0.16,0.12){13}{\line(1,0){0.16}}
\multiput(25.11,95.13)(0.18,0.12){11}{\line(1,0){0.18}}
\multiput(27.13,96.44)(0.20,0.11){10}{\line(1,0){0.20}}
\multiput(29.15,97.57)(0.25,0.12){8}{\line(1,0){0.25}}
\multiput(31.17,98.50)(0.29,0.11){7}{\line(1,0){0.29}}
\multiput(33.19,99.24)(0.40,0.11){5}{\line(1,0){0.40}}
\multiput(35.21,99.79)(0.67,0.12){3}{\line(1,0){0.67}}
\multiput(37.23,100.15)(1.01,0.08){2}{\line(1,0){1.01}}
\put(39.25,100.32){\line(1,0){2.02}}
\multiput(41.27,100.30)(1.01,-0.11){2}{\line(1,0){1.01}}
\multiput(43.30,100.08)(0.51,-0.10){4}{\line(1,0){0.51}}
\multiput(45.32,99.67)(0.34,-0.10){6}{\line(1,0){0.34}}
\multiput(47.34,99.07)(0.29,-0.11){7}{\line(1,0){0.29}}
\multiput(49.36,98.28)(0.22,-0.11){9}{\line(1,0){0.22}}
\multiput(51.38,97.29)(0.20,-0.12){10}{\line(1,0){0.20}}
\multiput(53.40,96.12)(0.17,-0.11){12}{\line(1,0){0.17}}
\multiput(55.42,94.75)(0.16,-0.12){13}{\line(1,0){0.16}}
\multiput(57.44,93.19)(0.13,-0.12){15}{\line(1,0){0.13}}
\multiput(59.46,91.44)(0.12,-0.11){17}{\line(1,0){0.12}}
\multiput(61.49,89.50)(0.12,-0.13){30}{\line(0,-1){0.13}}
\multiput(15.00,85.66)(0.12,-0.14){17}{\line(0,-1){0.14}}
\multiput(17.01,83.35)(0.12,-0.12){17}{\line(0,-1){0.12}}
\multiput(19.02,81.23)(0.12,-0.11){17}{\line(1,0){0.12}}
\multiput(21.02,79.31)(0.13,-0.12){15}{\line(1,0){0.13}}
\multiput(23.03,77.57)(0.15,-0.12){13}{\line(1,0){0.15}}
\multiput(25.04,76.03)(0.17,-0.11){12}{\line(1,0){0.17}}
\multiput(27.05,74.69)(0.20,-0.12){10}{\line(1,0){0.20}}
\multiput(29.06,73.54)(0.25,-0.12){8}{\line(1,0){0.25}}
\multiput(31.06,72.58)(0.29,-0.11){7}{\line(1,0){0.29}}
\multiput(33.07,71.82)(0.40,-0.11){5}{\line(1,0){0.40}}
\multiput(35.08,71.24)(0.50,-0.09){4}{\line(1,0){0.50}}
\multiput(37.09,70.87)(1.00,-0.09){2}{\line(1,0){1.00}}
\put(39.10,70.68){\line(1,0){2.01}}
\multiput(41.10,70.69)(1.00,0.10){2}{\line(1,0){1.00}}
\multiput(43.11,70.90)(0.50,0.10){4}{\line(1,0){0.50}}
\multiput(45.12,71.29)(0.40,0.12){5}{\line(1,0){0.40}}
\multiput(47.13,71.88)(0.29,0.11){7}{\line(1,0){0.29}}
\multiput(49.14,72.67)(0.22,0.11){9}{\line(1,0){0.22}}
\multiput(51.14,73.64)(0.20,0.12){10}{\line(1,0){0.20}}
\multiput(53.15,74.81)(0.17,0.11){12}{\line(1,0){0.17}}
\multiput(55.16,76.18)(0.15,0.12){13}{\line(1,0){0.15}}
\multiput(57.17,77.74)(0.13,0.12){15}{\line(1,0){0.13}}
\multiput(59.18,79.49)(0.12,0.11){17}{\line(1,0){0.12}}
\multiput(61.18,81.43)(0.12,0.13){17}{\line(0,1){0.13}}
\multiput(63.19,83.57)(0.11,0.13){16}{\line(0,1){0.13}}
\put(15.00,85.66){\circle*{3.33}}
\put(65.00,85.66){\circle*{3.33}}
\put(65.00,84.33){\line(1,-1){12.00}}
\put(65.00,86.33){\line(1,-1){12.33}}
\put(14.67,86.33){\line(-1,-1){11.33}}
\put(15.00,84.66){\line(-1,-1){11.33}}
\multiput(105.00,92.33)(-0.23,0.11){8}{\line(-1,0){0.23}}
\multiput(103.13,93.24)(-0.11,0.13){7}{\line(0,1){0.13}}
\multiput(102.36,94.15)(0.11,0.30){3}{\line(0,1){0.30}}
\multiput(102.69,95.06)(0.21,0.12){11}{\line(1,0){0.21}}
\multiput(105.00,96.33)(0.23,0.11){8}{\line(1,0){0.23}}
\multiput(106.87,97.24)(0.11,0.13){7}{\line(0,1){0.13}}
\multiput(107.64,98.15)(-0.11,0.30){3}{\line(0,1){0.30}}
\multiput(107.31,99.06)(-0.21,0.12){11}{\line(-1,0){0.21}}
\multiput(105.00,100.33)(-0.23,0.11){8}{\line(-1,0){0.23}}
\multiput(103.13,101.24)(-0.11,0.13){7}{\line(0,1){0.13}}
\multiput(102.36,102.15)(0.11,0.30){3}{\line(0,1){0.30}}
\multiput(102.69,103.06)(0.21,0.12){11}{\line(1,0){0.21}}
\multiput(105.00,104.33)(0.23,0.11){8}{\line(1,0){0.23}}
\multiput(106.87,105.24)(0.11,0.13){7}{\line(0,1){0.13}}
\multiput(107.64,106.15)(-0.11,0.30){3}{\line(0,1){0.30}}
\multiput(107.31,107.06)(-0.21,0.12){11}{\line(-1,0){0.21}}
\multiput(105.00,108.33)(-0.23,0.11){8}{\line(-1,0){0.23}}
\multiput(103.13,109.24)(-0.11,0.13){7}{\line(0,1){0.13}}
\multiput(102.36,110.15)(0.11,0.30){3}{\line(0,1){0.30}}
\multiput(102.69,111.06)(0.21,0.12){11}{\line(1,0){0.21}}
\multiput(105.00,112.33)(0.23,0.11){8}{\line(1,0){0.23}}
\multiput(106.87,113.24)(0.11,0.13){7}{\line(0,1){0.13}}
\multiput(107.64,114.15)(-0.11,0.30){3}{\line(0,1){0.30}}
\multiput(107.31,115.06)(-0.21,0.12){11}{\line(-1,0){0.21}}
\multiput(135.00,95.33)(-0.23,0.11){8}{\line(-1,0){0.23}}
\multiput(133.13,96.24)(-0.11,0.13){7}{\line(0,1){0.13}}
\multiput(132.36,97.15)(0.11,0.30){3}{\line(0,1){0.30}}
\multiput(132.69,98.06)(0.21,0.12){11}{\line(1,0){0.21}}
\multiput(135.00,99.33)(0.23,0.11){8}{\line(1,0){0.23}}
\multiput(136.87,100.24)(0.11,0.13){7}{\line(0,1){0.13}}
\multiput(137.64,101.15)(-0.11,0.30){3}{\line(0,1){0.30}}
\multiput(137.31,102.06)(-0.21,0.12){11}{\line(-1,0){0.21}}
\multiput(135.00,103.33)(-0.23,0.11){8}{\line(-1,0){0.23}}
\multiput(133.13,104.24)(-0.11,0.13){7}{\line(0,1){0.13}}
\multiput(132.36,105.15)(0.11,0.30){3}{\line(0,1){0.30}}
\multiput(132.69,106.06)(0.21,0.12){11}{\line(1,0){0.21}}
\multiput(135.00,107.33)(0.23,0.11){8}{\line(1,0){0.23}}
\multiput(136.87,108.24)(0.11,0.13){7}{\line(0,1){0.13}}
\multiput(137.64,109.15)(-0.11,0.30){3}{\line(0,1){0.30}}
\multiput(137.31,110.06)(-0.21,0.12){11}{\line(-1,0){0.21}}
\multiput(135.00,111.33)(-0.23,0.11){8}{\line(-1,0){0.23}}
\multiput(133.13,112.24)(-0.11,0.13){7}{\line(0,1){0.13}}
\multiput(132.36,113.15)(0.11,0.30){3}{\line(0,1){0.30}}
\multiput(132.69,114.06)(0.21,0.12){11}{\line(1,0){0.21}}
\multiput(135.00,115.33)(0.23,0.11){8}{\line(1,0){0.23}}
\multiput(136.87,116.24)(0.11,0.13){7}{\line(0,1){0.13}}
\multiput(137.64,117.15)(-0.11,0.30){3}{\line(0,1){0.30}}
\multiput(137.31,118.06)(-0.21,0.12){11}{\line(-1,0){0.21}}
\multiput(95.00,85.66)(0.12,0.13){17}{\line(0,1){0.13}}
\multiput(97.02,87.94)(0.12,0.12){17}{\line(0,1){0.12}}
\multiput(99.04,90.02)(0.13,0.12){16}{\line(1,0){0.13}}
\multiput(101.06,91.92)(0.13,0.11){15}{\line(1,0){0.13}}
\multiput(103.08,93.62)(0.16,0.12){13}{\line(1,0){0.16}}
\multiput(105.11,95.13)(0.18,0.12){11}{\line(1,0){0.18}}
\multiput(107.13,96.44)(0.20,0.11){10}{\line(1,0){0.20}}
\multiput(109.15,97.57)(0.25,0.12){8}{\line(1,0){0.25}}
\multiput(111.17,98.50)(0.29,0.11){7}{\line(1,0){0.29}}
\multiput(113.19,99.24)(0.40,0.11){5}{\line(1,0){0.40}}
\multiput(115.21,99.79)(0.67,0.12){3}{\line(1,0){0.67}}
\multiput(117.23,100.15)(1.01,0.08){2}{\line(1,0){1.01}}
\put(119.25,100.32){\line(1,0){2.02}}
\multiput(121.27,100.30)(1.01,-0.11){2}{\line(1,0){1.01}}
\multiput(123.30,100.08)(0.51,-0.10){4}{\line(1,0){0.51}}
\multiput(125.32,99.67)(0.34,-0.10){6}{\line(1,0){0.34}}
\multiput(127.34,99.07)(0.29,-0.11){7}{\line(1,0){0.29}}
\multiput(129.36,98.28)(0.22,-0.11){9}{\line(1,0){0.22}}
\multiput(131.38,97.29)(0.20,-0.12){10}{\line(1,0){0.20}}
\multiput(133.40,96.12)(0.17,-0.11){12}{\line(1,0){0.17}}
\multiput(135.42,94.75)(0.16,-0.12){13}{\line(1,0){0.16}}
\multiput(137.44,93.19)(0.13,-0.12){15}{\line(1,0){0.13}}
\multiput(139.46,91.44)(0.12,-0.11){17}{\line(1,0){0.12}}
\multiput(141.49,89.50)(0.12,-0.13){30}{\line(0,-1){0.13}}
\multiput(95.00,85.66)(0.12,-0.14){17}{\line(0,-1){0.14}}
\multiput(97.01,83.35)(0.12,-0.12){17}{\line(0,-1){0.12}}
\multiput(99.02,81.23)(0.12,-0.11){17}{\line(1,0){0.12}}
\multiput(101.02,79.31)(0.13,-0.12){15}{\line(1,0){0.13}}
\multiput(103.03,77.57)(0.15,-0.12){13}{\line(1,0){0.15}}
\multiput(105.04,76.03)(0.17,-0.11){12}{\line(1,0){0.17}}
\multiput(107.05,74.69)(0.20,-0.12){10}{\line(1,0){0.20}}
\multiput(109.06,73.54)(0.25,-0.12){8}{\line(1,0){0.25}}
\multiput(111.06,72.58)(0.29,-0.11){7}{\line(1,0){0.29}}
\multiput(113.07,71.82)(0.40,-0.11){5}{\line(1,0){0.40}}
\multiput(115.08,71.24)(0.50,-0.09){4}{\line(1,0){0.50}}
\multiput(117.09,70.87)(1.00,-0.09){2}{\line(1,0){1.00}}
\put(119.10,70.68){\line(1,0){2.01}}
\multiput(121.10,70.69)(1.00,0.10){2}{\line(1,0){1.00}}
\multiput(123.11,70.90)(0.50,0.10){4}{\line(1,0){0.50}}
\multiput(125.12,71.29)(0.40,0.12){5}{\line(1,0){0.40}}
\multiput(127.13,71.88)(0.29,0.11){7}{\line(1,0){0.29}}
\multiput(129.14,72.67)(0.22,0.11){9}{\line(1,0){0.22}}
\multiput(131.14,73.64)(0.20,0.12){10}{\line(1,0){0.20}}
\multiput(133.15,74.81)(0.17,0.11){12}{\line(1,0){0.17}}
\multiput(135.16,76.18)(0.15,0.12){13}{\line(1,0){0.15}}
\multiput(137.17,77.74)(0.13,0.12){15}{\line(1,0){0.13}}
\multiput(139.18,79.49)(0.12,0.11){17}{\line(1,0){0.12}}
\multiput(141.18,81.43)(0.12,0.13){17}{\line(0,1){0.13}}
\multiput(143.19,83.57)(0.11,0.13){16}{\line(0,1){0.13}}
\put(95.00,85.66){\circle*{3.33}}
\put(145.00,85.66){\circle*{3.33}}
\put(145.00,84.33){\line(1,-1){12.00}}
\put(145.00,86.33){\line(1,-1){12.33}}
\put(94.67,86.33){\line(-1,-1){11.33}}
\put(95.00,84.66){\line(-1,-1){11.33}}
\multiput(105.00,76.00)(-0.23,0.11){8}{\line(-1,0){0.23}}
\multiput(103.13,76.91)(-0.11,0.13){7}{\line(0,1){0.13}}
\multiput(102.36,77.82)(0.11,0.30){3}{\line(0,1){0.30}}
\multiput(102.69,78.73)(0.21,0.12){11}{\line(1,0){0.21}}
\multiput(105.00,80.00)(0.23,0.11){8}{\line(1,0){0.23}}
\multiput(106.87,80.91)(0.11,0.13){7}{\line(0,1){0.13}}
\multiput(107.64,81.82)(-0.11,0.30){3}{\line(0,1){0.30}}
\multiput(107.31,82.73)(-0.21,0.12){11}{\line(-1,0){0.21}}
\multiput(105.00,84.00)(-0.23,0.11){8}{\line(-1,0){0.23}}
\multiput(103.13,84.91)(-0.11,0.13){7}{\line(0,1){0.13}}
\multiput(102.36,85.82)(0.11,0.30){3}{\line(0,1){0.30}}
\multiput(102.69,86.73)(0.21,0.12){11}{\line(1,0){0.21}}
\multiput(105.00,88.00)(0.23,0.11){8}{\line(1,0){0.23}}
\multiput(106.87,88.91)(0.11,0.13){7}{\line(0,1){0.13}}
\multiput(107.64,89.82)(-0.11,0.30){3}{\line(0,1){0.30}}
\multiput(107.31,90.73)(-0.21,0.12){11}{\line(-1,0){0.21}}
\put(40.00,55.66){\makebox(0,0)[cc]{(a)}}
\put(120.00,55.66){\makebox(0,0)[cc]{(b)}}
\end{picture}
\vskip -3cm
{\bf Fig.1} \ Photon absorption graphs.
\label{fig1}
\vskip 1.5cm
\end{figure}

Let us calculate $C(q^2_\bot)$ in the relativistic constituent quark
framework. Since the photon can be absorbed by each of the constituents
of the pion, two kind of terms arises, as shown in 
fig.\ref{fig1}. One
corresponds to the direct term (D), which has the same quark absorbing
and emitting the photon.  The other one, the exchange term (E), has the
photon absorbed by one quark and emitted by the other in its hermitian
conjugate. 
We assume that the quarks could be treated as if they are free in the 
final state, however in reality the quarks are confined. The 
validity of such hypothesis was discussed in detail by 
Jaffe\cite{jaffe} in the case of form-factors, and he concluded that
in many cases the physics is dominated by aspects of the wave function
not directly related with the confinement. 
We consider in our approach that, for medium $q^2$ range, the 
characteristic distances between the quarks in the final 
states are below the confinement scale.

Then, the electromagnetic tensor is written as

\begin{eqnarray}
W^{\mu\nu}=W_D^{\mu\nu}+W_E^{\mu\nu} \ .
\end{eqnarray}
In the Bjorken limit, $q_\bot^2\ \rightarrow \ \infty$, just the direct
term $W_D$ survives, and the exchange term $W_E$ is nonzero only as
high twist correction. Correspondingly, the integral $C$, as given in
Eq.(\ref{c}), is expressed as the sum of a direct and an exchange term:

\begin{equation}
C(q^2_\perp) = \ C_{D}(q^2_\perp)\ + \ C_{E}(q^2_\perp)
\label{eq:cde}
\end{equation}
The coupling of the pion with quarks is given by the effective
Lagrangian with vertex ${\it L}^{eff}_{\pi \to q\bar q} =
(M/f_\pi)g_\pi(p^2,(p_1-p_2)^2)\overline q \gamma^5{{\bf \tau}} q$
defining the soft transition amplitude of the pion into quark -
antiquark pair $\pi (p) \rightarrow \overline q(p_1) q(p_2)$, where $M$
is the constituent quark mass, ${\bf \tau}$ the isospin matrices, and
$f_\pi=93\ MeV$.

>From the direct component of the pion electromagnetic tensor 
(fig.\ref{fig1}a),
using Eq.(\ref{c}), at $q^+=0$ and fixed ${\bf q}_\bot^2$ ($q^2 = -{\bf
q}_\bot^2$), we obtain:
{\small
\begin{eqnarray} 
&&C_D(q^2_\perp) = I^\alpha_D \frac{N_c}{2\pi m_\pi}
\frac{M^2} {f^2_\pi}\int dq^-  \frac{d^4k d^4k'}{(4\pi)^2}k'^+
g^2_\pi\left( \left[k-\frac{p}{2}\right]^2\right)
\delta^4(k'-k+p_\pi+q) 
\nonumber \\ &&
\times \delta(k^2 -M^2) \delta(k'^2 - M^2) 
\frac{tr \left[\gamma^+ (\rlap\slash{k} -
\rlap\slash{p}_\pi + M)
\gamma^5 (\rlap\slash{k} + M)
\gamma^5 (\rlap\slash{k} - \rlap\slash{p}_\pi + M)\right]}
{\left[(k - p_\pi)^2 - M^2\right]^2},
\label{cd1}
\end{eqnarray} }
where $N_c $ is the number of colors, $k$ is the 4-momentum of the
spectator quark, $ k' = k - p_\pi - q$, and $\gamma^+=
\gamma^0+\gamma^3$.  The trace over isospin space, considering the
charges of quark and antiquark, gives $I_D^\alpha$:

\[I_D^\alpha = Tr\left(QQ\tau^\alpha{\tau^\alpha}^\dagger \right) +
Tr\left(\bar Q \bar Q {\tau^\alpha}^\dagger \tau^\alpha \right) =
\frac{5}{9} \;
{for \; \; } \; \alpha = \pi^+, \; \pi^-, \; \pi^o, \]
where the charge matrices of quark and antiquark are
$Q = - \bar Q = (1/6 + \tau_z/2)$.

The integrations over the four momentum $k'$ and over the light cone
variables $q^-$ and $k^-$, in Eq.~(\ref{cd1}), provide the result:

\begin{eqnarray}
C_D(q^2_\perp) \ = \ I_D^\alpha \frac{2 N_c }{(2\pi)^3} \
\frac{M^2}{f^2_\pi} \int^1_0 dx \int \frac{d^2k_\perp}{x (1 - x)}
\frac{M^2_0} {\left(M^2_0-m^2_\pi \right)^2 }g^2_\pi(M_0^2) \ ,
\label{cd2}
\end{eqnarray}
where the momentum fraction $x = k^+/m_\pi$ is introduced and the
invariant mass of the $q\overline q$ system is given by

\begin{eqnarray}
M^2_0(x,{{\bf k}_\perp})=\frac{k^2_\perp+M^2}{x(1-x)} \ .
\label{eq:m02}
\end{eqnarray}
>From Eq. (\ref{cd2}) it is easy to see that $C_D(q^2_\bot)$ is
proportional to the normalization factor of the pion elastic
electromagnetic form-factor
\cite{teren,BETI,tg} and the pion deep inelastic structure function
\cite{RR}.  Thus, we have

\[C_D(q^2_\perp) \ =\ \frac{5}{9} ,\]
that is the sum of the valence quark number weighted by the squared
charges.

We observe that, at few GeV, the 
sum rule should approach the deep inelastic sum rule,
which means that it becomes proportional to $\int dx (q(x) + \bar q(x))$,
which diverges, signaling the presence of an infinite number
of partons. In our picture these partons are present in the constituent quark. 
However, we exclude  the direct term in our approach, by considering 
the difference between 
the sum rules for charged and non-charged pions, introducing the 
quark - antiquark density.

Next, in analogous manner, we evaluate the exchange component
$C_E(q^2_\perp)$ (fig.\ref{fig1}b):
{\small \begin{equation}
C_E(q^2_\perp)  =I_E^\alpha
\frac{2 N_c}{(2\pi)^3} \ \frac{M^2}{f^2_\pi}
\int^1_0 dx \int \frac{d^2k_\perp}{x (1 - x)}
\frac{\left( M^2_0  - \frac{ {\bf k}_\perp {\bf .
q}_\perp}{[x(1-x)]}\right) g_\pi(M_0^2)g_\pi(M'^2_0)}
{\left(M'^2_0-m^2_\pi \right) \left(M^2_0-m^2_\pi\right)}, 
\label{eq:ce2}
\end{equation} }
where $M^2_0 \equiv M^2_0(x,{\bf k}_\perp)$ and
$M'^2_0 \equiv M^2_0(x,{\bf k}_\perp -{\bf q}_\perp)$ are
given by Eq.(\ref{eq:m02}),
and

\begin{eqnarray}
I_E^\alpha = 2 Tr \left( Q \tau^\alpha \bar Q {\tau^\alpha}^\dagger
\right) &=& - 5/9 \; {for} \; \; \pi^0 ;
\nonumber \\
&=& + 4/9 \; {for} \; \; \pi^\pm .
\end{eqnarray}

>From Eq.(\ref{eq:ce2}), we can define the quark - antiquark density
$C_{q\bar q}(q^2_\perp)$,
\begin{equation}
C_{q\bar q}(q^2_\perp) = \frac{C_E(q^2_\perp)}{I_E^\alpha}.
\label{eq:cpi}
\end{equation}
We can see now that $C_E(q_\perp^2)$ is proportional to the
normalization factor only at $q_\perp^2 = 0$, such that 
$C_{q\bar q}(0) \ = \ 1.$

To define the correlation  function let us separate out the elastic
contribution of the sum rule, Eq.(\ref{c}). The matrix elements of the
electromagnetic current between pion states are expressed as
\begin{equation}
\langle\pi(p')| J^\gamma_\mu (0) |\pi(p)\rangle  =
(p_\mu+p'_\mu) F_\pi (q^2),
\label{Jsc}\end{equation}
where $F_\pi(q^2)$ is the electromagnetic form factor of the pion
normalized at the origin: $F_\pi(0)=1$.  It is easy to see, from the
definition in Eq.(\ref{wgam}), that in the elastic limit the
contribution to the sum rule will be
\begin{equation}
C_{elastic}(q^2_\perp) \ = \ F_\pi ^2(q^2_\bot).
\label{Welast}\end{equation}
We define the correlation function characterizing the  deviation of the
exchange sum from the elastic contribution as
\begin{equation}
C_{corr}(q^2_\perp) \ = C_{q\bar q}(q^2_\perp) -
C_{elastic}(q^2_\perp).
\label{Ccorr}
\end{equation}
>From this definition it follows that the absolute value of the
correlation function $C_{corr}(q^2_\perp) $ is zero at $q^2_\perp = 0$
and as $q^2_\perp \to \infty$, and has an extremum at $q^2_\perp =\bar
q^2_\perp$. \ The maximum of the absolute value of the correlation
function defines the quark - antiquark correlation length:
\begin{equation}
l_{corr}  =1/ \sqrt{\bar q^2_\perp}.
\label{lcorr}
\end{equation}
Then, following refs. \cite{wale},
the total sum rule can be expressed as the sum of the elastic
contribution ($C_{elastic}$), inelastic contribution in the absence of
correlations ($C_D - (5/9) F_\pi^2$), and inelastic contribution in
presence of correlations, $C_{corr}$:
\begin{eqnarray}
C^{\pm}(q^2_\perp) =& F_\pi^2(q^2_\perp) +
			\frac{5}{9}[1- F_\pi ^2(q^2_\perp)] +
			\frac{4}{9} C_{corr}(q^2_\perp) &=
			\frac{5}{9}+\frac{4}{9}C_{q\bar q}(q^2_\perp),
\nonumber \\
C^{0}(q^2_\perp) =& \frac{5}{9}\{[1- F_\pi ^2(q^2_\perp)]  -
		       C_{corr}(q^2_\perp) \} &=
		       \frac{5}{9}-\frac{5}{9}C_{q\bar q}(q^2_\perp).
\label{SRtot}
\end{eqnarray}
Here we have to note that in our calculations we didn't take into
account the Pomeron exchange contribution. To exclude it we consider
the difference between the total sum rules for charged and non-charged
pions which directly defines the quark - antiquark density, $C_{q\bar
q}$:
\begin{equation}
C^{\pm}(q^2_\perp) -  C^{0}(q^2_\perp)=
F_\pi^2(q^2_\perp) + C_{corr}(q^2_\perp) =  C_{q\bar q}(q^2_\perp).
\label{Gsr}\end{equation}
This subtraction removes the contribution of the 
direct term, which survives in the deep-inelastic limit. Experimentally
the direct term is divergent, but such subtraction turns Eq.(\ref{Gsr})
finite.

The cross-section for inelastic electron scattering  on the pion, 
in the medium range $q^2$, where the resonances dominate, allows to
address experimentally the correlation length 
through $C_{corr}(q^2_\perp)$. This last quantity
comes from the difference between the experimental
structure functions $W_2$ for the charged and uncharged pion, 
integrated in the transferred energy, at a fixed $q^2$,
 \begin{equation}
C_{corr}(q^2_\perp) =\frac{1}{e^2} \int^\infty_{\nu_0} 
d\nu (W^{\pm}_2( q^2,\nu)-W^{0}_2( q^2,\nu)) 
 \ , 
 \end{equation}
 where $\nu_0$ is the inelastic threshold for the process.
 A similar procedure has been applied 
for the nucleon, where the difference between the correlation functions 
of the proton and the neutron has been obtained from the inelastic 
electron scattering data \cite{Schmidt}.

The light-front pion wave-function can be introduced by modifying the
vertex as discussed in refs. \cite{teren,BETI,tg}.
In this scheme the composite pion has the correct quantum numbers,
which is equivalent to constructing the pion wave-function as in
ref.~\cite{chupi88}.  The light-front pion wave-function in terms of
the relative coordinate, can be introduced as in
ref.~\cite{tg}\footnote{In this reference, $g_\pi(M_0^2) = 1$.}, by the
following

\begin{eqnarray}
\Phi_\pi(x, {\bf k}_\perp)\ = \frac{1}{\pi^\frac32}\frac{M}{f_\pi}
\frac{\sqrt{M_0 N_c}}{M^2_0-m^2_\pi }g_\pi(M_0^2).
 \label{eq:wf}
\end{eqnarray}

Substituting in Eq. (\ref{eq:ce2}) the mass denominator by the
bound-state wave-function given by Eq.(\ref{eq:wf}), 
we have the result for the quark-antiquark density function in 
the pion:
\begin{eqnarray}
C_{q\bar q}(q^2_\perp)  = 
\int^1_0 dx \int \frac{d^2k_\perp}{4x (1 - x)}
\left[M_0^2- \frac{{\bf k}_\perp . {\bf q}_\perp}{x(1-x)}\right]
\frac{ \Phi_\pi(x, {\bf k}_\perp)\ 
\Phi_\pi(x, {\bf k}_\perp-{\bf q}_\perp)}{\sqrt{M_0 M'_0}}
\ .
\label{eq:ce3}
\end{eqnarray}
With this choice of a phenomenological wave-function, we have the usual
non-relativistic normalization, that is obtained at ${\bf q}_\perp$ =
0, according to the context of the Hamiltonian Front Form of the
dynamics \cite{chupi88,chung88}:

\begin{equation}
\int d^3k \left[\Phi_\pi( {\bf k})\right]^2 \ = \ 1 .
\label{eq:norm}
\end{equation}
Here and in the following expressions, we use a dual notation, when
writing our functions in terms of the instant form variables and in
terms of the light-cone variables, such that $$\Phi ({\bf k}) \equiv
\Phi (x, {\bf k}_\perp).$$ The third component of the momentum is given
in terms of $x$ and ${\bf k}_\perp$ by \cite{chupi88}

\begin{equation}
 k_z=\left( x-\frac12\right)\sqrt{ \frac{k_\perp^2+M^2}{x(1-x)} } =
\left( x-\frac{1}{2}\right) M_0 \ ,
\end{equation}
and the Jacobian of the transformation between $(x,{\bf k}_\perp)$ and
${\bf k} $ is
\begin{equation}
\frac{\partial (x,{\bf k}_\perp)}{ \partial (k_z,{\bf k}_\perp)} \ = \
4 \left[ \frac{ [x(1-x)]^3}{k^2_\perp +M^2} \right]^{1/2} \ =
\ \frac{4x(1-x)}{M_0} .
\end{equation}

The transverse momentum, in the argument of the light-front
wave-functions in Eq.(\ref{eq:ce3}), is given in the pion center of
mass. The transverse photon momentum is subtracted from the center of
mass momentum of one of the quarks, as a consequence of the absorption
of the photon by the quark (antiquark) and the subsequent emission by
the antiquark (quark).

In the non-relativistic limit $(M\rightarrow \infty)$, $C_{q\bar q}
(q^2_\perp)$ reduces to the Fourier transform of the two-body density,
which appears in the Coulomb sum-rule\cite{wale}, and it is given by:
\begin{eqnarray}
C^{NR}_{q\bar q} (q^2_\perp) \ = \ \int d^3k
\Phi_\pi({\bf k})\ \Phi_\pi({\bf k} -{\bf q})\
\ .
\label{eq:cenr}
\end{eqnarray}

For completeness, we present below the  expressions of the pion charge
form-factor, $F_\pi(q^2_\perp)$, and weak decay constant, $f_\pi$,
using the light-front wave-function \cite{teren,BETI,tg}:
\begin{eqnarray}
F_{\pi}(q^2_\perp) = 
\int^1_0 dx \int \frac{d^2k_\perp}{4x (1 - x)}
\left[M_0^2+ \frac{{\bf k}_\perp . {\bf q}_\perp}{x}\right]
\frac{\Phi_\pi(x, {\bf k}_\perp)\ \Phi_\pi(x, {\bf k}'_\perp )}
{\sqrt{M_0 \tilde M_0}}
\ ,
\label{eq:ff}
\end{eqnarray}
where
${\bf k}'_\perp \equiv {\bf k}_\perp+(1-x){\bf q}_\perp $
and
$\tilde M_0 \equiv M_0(x, {\bf k}'_\perp )$.

We have to emphasize that, inspite of similarity in the form of the
expressions given in Eqs.(\ref{eq:ce3}) and (\ref{eq:ff}) (just change
$- q_\perp \to (1-x){\bf q}_\perp$), they have very different physical
interpretations.  In the expression for $F_{\pi}(q^2_\perp)$, one of
the pion is boosted such that it absorbs all the photon momentum in the
form factor, whereas in the exchange term $C_{E}(q^2_\perp)$ the wave
functions are calculated in the rest frame of pion. Covariance under
kinematical boost guarantee that we can obtain the boosted wave
function from the wave function in the center of mass frame (see
ref.~\cite{tob2} and references therein).

With the definition given by Eq.(\ref{eq:wf}), the weak decay constant
is given by:
\begin{eqnarray}
f_\pi= \frac{M\sqrt{N_c}}{4 \pi^{\frac32}}
\int \frac{dx d^2k_\perp}{x(1-x)} \frac{\Phi_\pi(x,k_\perp)}{\sqrt{M_0}}.
\label{eq:fpi}
\end{eqnarray}

Let us make some predictions of the pion quark - antiquark density and
correlation function, by performing numerical calculations with four
models of light-front wave-functions: i) instanton, ii) Gaussian, iii)
hydrogen atom and iv) the model wave-function of ref.~\cite{god}.  A
recent discussion about light-front pion wave-functions can be also
found in ref.~\cite{shakin}. / Except for the total elastic
contribution, as one can see from Eq.(\ref{Ccorr}), the correlation
function is closely related to the quark - antiquark density.

 The hadron wave functions are defined by low energy quark dynamics.
Within the realistic QCD vacuum approach, like QCD sum rules or
instanton liquid model, the hadrons are considered as low energy
excitations of nonperturbative QCD vacuum. As it has been shown by 't
Hooft~\cite{hooft}, for small size instanton the interaction generated
by instanton - antiinstanton configurations induces a chirally
invariant four - quark interaction, whose contributions to the
Lagrangian is of the form \begin{equation} G \left[ (\bar \Psi \Psi)^2
- (\bar \Psi \gamma_5 {\bf \tau} \Psi)^2 \right], \label{qfield}
\end{equation} where $\Psi$ is the quark field and $G$ is the
interaction constant.

 The quark then acquires a momentum dependent mass $M\cdot g(p^2)$
(where M is the constituent quark mass and $g(0)=1$), via the
Nambu-Jona-Lasinio mechanism \cite{NJL}, and in addition bound states
appear in the pseudo scalar channel of $q\bar q$ system. From quark -
antiquark scattering in the field of instanton, the non-local vertex
function $g(p^2)$ is derived, with $p^2$ being invariant mass of quark
- antiquark system \cite{doro,RR}.

The instanton inspired wave function is defined by the vertex function
\begin{equation}
g_{inst} (x,k_\perp ) = exp{ \left[\frac{-\sqrt{\lambda}}{2}
M_0(x,k_\perp )\right]},
\end{equation}
such that, within the normalization given by Eq.(\ref{eq:norm}),
we have
\begin{equation}
\Phi_{inst}({\bf k}) = N_{inst} \frac{\left[4 ({\bf
k}^2+M^2)\right]^{1/4}} {\left[4({\bf k}^2+M^2)-m_\pi^2 \right]}
exp{\left(-\sqrt{ \lambda ({\bf k}^2+M^2)} \right) },
\label{eq:wfins}
\end{equation}
where $N_{inst}$ is the normalization factor.
In terms of the invariant $q\bar q$ mass we have
\begin{equation}
\Phi_{inst}(x,{\bf k}_\perp) = N_{inst}
\frac{\sqrt M_0}{\left[M_0^2-m_\pi^2 \right]}
exp{\left(-\sqrt{\lambda} \frac{M_0}{2}  \right)}.
\label{eq:wf2}
\end{equation}

The Gaussian wave-function is given by

\begin{equation}
\Phi_\pi ({\bf k}) = \left(\frac{8 r^2_{NR}}{3\pi}\right)^{3/4}
exp\left(-\frac{4}{3} r^2_{NR} {\bf k}^2 \right) \ ;
\end{equation}
and the hydrogen atom wave-function by

\begin{equation}
\Phi_\pi({\bf k}) = \frac{1}{2\pi}\left(\frac{\sqrt 3}
{r_{NR}}\right)^{5/2}
\left( \frac{3}{4} r_{NR}^{-2}   + {\bf k}^2\right)^{-2} \ .
\end{equation}
In these last two wave-functions, $r_{NR}$ is the scale defining the
size properties of the wave function and the constituent quark mass we
fixed at the value of 220 MeV, as in the model of ref.\cite{god}.

The electromagnetic pion radius is given by Eq.(\ref{eq:ff}), through
the expression
\begin{equation}
r_\pi \ = \ \sqrt {- 6 
\left[\frac{dF_\pi}{dq^2_\perp}\right] }_{q^2_\perp  =0};
\label{eq:rpi}
\end{equation}
and the quark - antiquark density radius $r_{q\bar q}$ is given by
Eq.(\ref{eq:ce3}):
\begin{equation}
r_{q\bar q} \ = \ \sqrt{- 6 
\left[\frac{d C_{q\bar q}(q^2_\perp)}
{dq^2_\perp}\right]}_{q^2_\perp  =0}.
\label{eq:rcor}
\end{equation}
In the same way, we have the correlation radius, defined from
 Eq.(\ref{Ccorr}):
\begin{equation}
r_{corr} \ = \ \sqrt{- 6 
\left[\frac{d C_{corr}(q^2_\perp)}{dq^2_\perp}\right]}
_{q^2_\perp  =0} = \sqrt{r^2_{q\bar q}-2r^2_\pi}.
\label{rwcor}
\end{equation}

It is well known that the square radius of the transverse space
correlation function can be related with the total photoproduction
cross section $\sigma_\gamma^T$. For the charged - neutral pion
difference one has
\begin{equation}
\frac{d}{dq^2_\perp}\left[C^{\pm}(q^2_\perp) -  
C^{0}(q^2_\perp)\right]|_{q^2_\perp =0}= 
-\frac{1}{3}\langle r^2_\pi\rangle +
\frac{1}{4\pi^2\alpha}\int\limits_{\nu_0}^{\infty}\frac{d\nu}{\nu}
(\sigma_{\gamma\pi^\pm}^T - \sigma_{\gamma\pi^0}^T),
\label{GSBsr}\end{equation}
where $\nu_0$ is the threshold for inelastic photon absorption. and
\begin{equation} -\frac{1}{m_\pi} \frac{dW_2(q^2,\nu)}{dq^2} =
\frac{\sigma_{\gamma\pi}^T} {4\pi^2\alpha\nu}.
\label{SigTot}\end{equation}

This sum rule has been first derived by Gerasimov \cite{GSBsr}.  In
order to obtain the estimation of the integral of the difference of the
cross sections, that appears in the left hand side of
Eq.~(\ref{GSBsr}), we can rewrite this sum rule in a form that it is
related with $r_{corr}$:
\begin{equation}
\frac{1}{4\pi^2\alpha}\int\limits_{\nu_0}^{\infty}\frac{d\nu}{\nu}
(\sigma_{\gamma\pi^\pm}^T - \sigma_{\gamma\pi^0}^T) =
-\frac{1}{6}\langle r^2_{corr}\rangle
\label{GSBsr1}\end{equation}
Analogous considerations in nucleon case has been done in 
refs.~\cite{Gott,Schmidt}.

\begin{figure}
\postscript{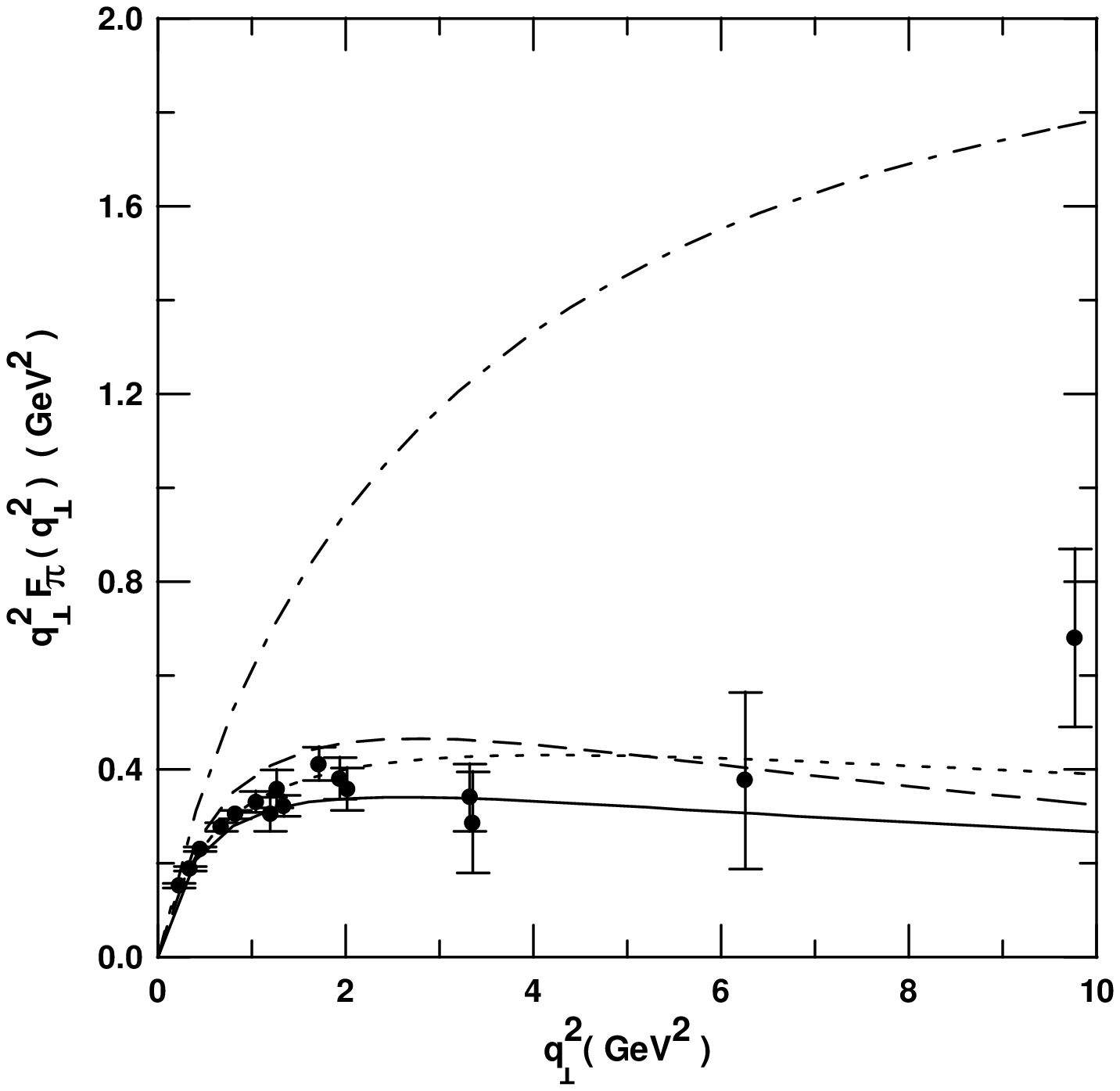}{1.0}
\vskip -3cm  
{\bf Fig.2} \ Pion form-factor for  $q^2_\perp \ <$ 10 GeV$^2$.
The four curves represent four models considered in this paper.
Godfrey and Isgur model (dot-dashed) uses the non-relativistic radius of
0.195 fm that does not fit $f_\pi$. The other three curves uses
parameters such that fit $f_\pi$ = 93 MeV. In the Instanton model
(dotted) we use $M$ = 200 MeV and $\lambda$ = 0.1153$/M$;
for the Hydrogen-atom model (solid) we use $M$ = 220 MeV and
$r_{NR}$= 0.456 fm; and for the Gaussian model (long-dashed)
 $M$ = 220 MeV and $r_{NR}$= 0.321 fm.
Experimental data from ref.\cite{exp}.
\label{fig2}
\vskip 0.5cm
\end{figure}

The wave function model given in ref. \cite{god} has a non-relativistic
pion radius of 0.195 fm and it gives for the electromagnetic pion
radius the value of 0.456 fm~\cite{tg}. For the same $r_{NR}$ the
Gaussian and the hydrogen-atom models give $r_{\pi}$ values of 0.476 fm
and 0.463 fm, respectively\cite{tg}. None of these three models were
able to describe the experimental electromagnetic pion radius of
$r_\pi^{exp}$ = 0.660 $\pm$ 0.024 fm~\cite{rexp}. \ We observe, in this
case, that the experimental values of $F_\pi(q^2_\perp)$~\cite{exp} are
not reproduced by the models, and in 
fig.\ref{fig2} this fact is represented by
the model of ref.~\cite{god}.


In order to have a model reasonable consistent with the experimental
form factor data, we choose to fit the experimental pion decay constant
$f_\pi$ = 93 MeV, since this also should produce a reasonable pion
radius \cite{tg}.  We obtain $r_{NR}$ = 0.321 fm and $r_\pi$ = 0.64 fm
for the Gaussian model, and  $r_{NR}$ = 0.456 fm and $r_\pi$ = 0.76 fm
for the hydrogen-atom model. With these parameters the calculated pion
form factor for both models are in good agreement with the experimental
data, as we observe in fig.\ref{fig2}.  Here we also show the results for the
instanton model, that fit $f_\pi$ with $M =$ 200 MeV and $\lambda =
0.1153/M^2$. The instanton model gives $r_\pi$ = 0.77 fm, and shows a
behaviour similar to the hydrogen atom model, with a good fitting of
the experimental data.

The calculated quark - antiquark density and correlation radius for the
different models we consider, using Eqs.(\ref{eq:rcor}) and
(\ref{rwcor}), are: $r_{q\bar q}$ = 1.12 fm and $r_{corr}$ = 0.66 fm
for the Gaussian model, $r_{q \bar q}$ = 1.37 fm and $r_{corr}$ = 0.85
fm for the Hydrogen-atom model, and $r_{q \bar q}$= 1.39 fm and
$r_{corr}$ = 0.86 fm for the Instanton model.  As we see, the quark -
antiquark density radius are near twice the pion radius, in agreement
with the nonrelativistic expectation.

\begin{figure}
\postscript{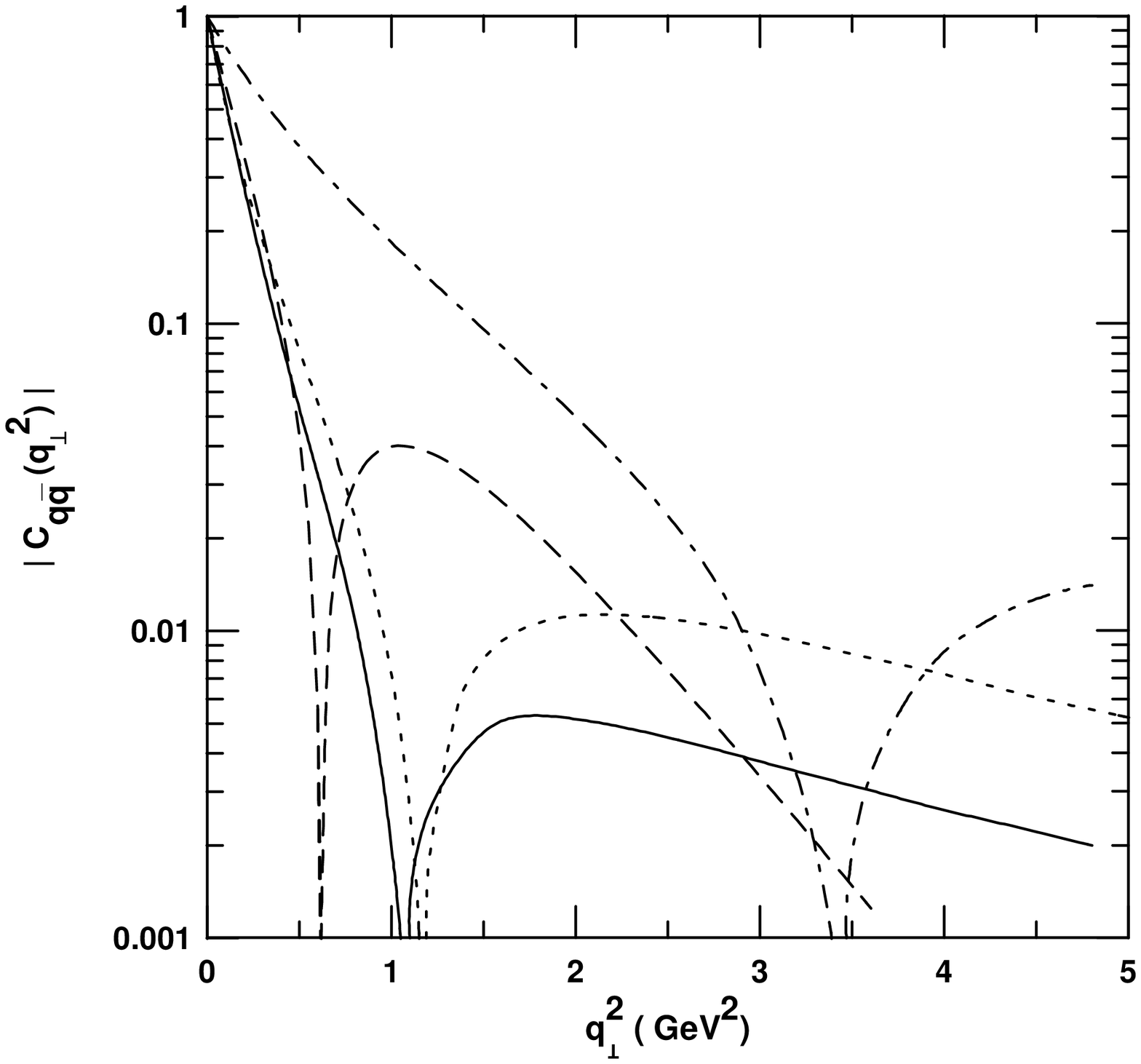}{0.6}
\vskip -2cm 
{\bf Fig.3} \ The absolute value of the quark - antiquark density 
in the pion, $ C_{q\bar q}(q^2_\perp) $ for $q^2_\perp \ <$ 5 GeV$^2$.
The curves represent the models with the parametrization and line
conventions as in fig.\ref{fig2}.
\vskip 1cm
\label{fig3}
\end{figure}

\begin{figure}
\postscript{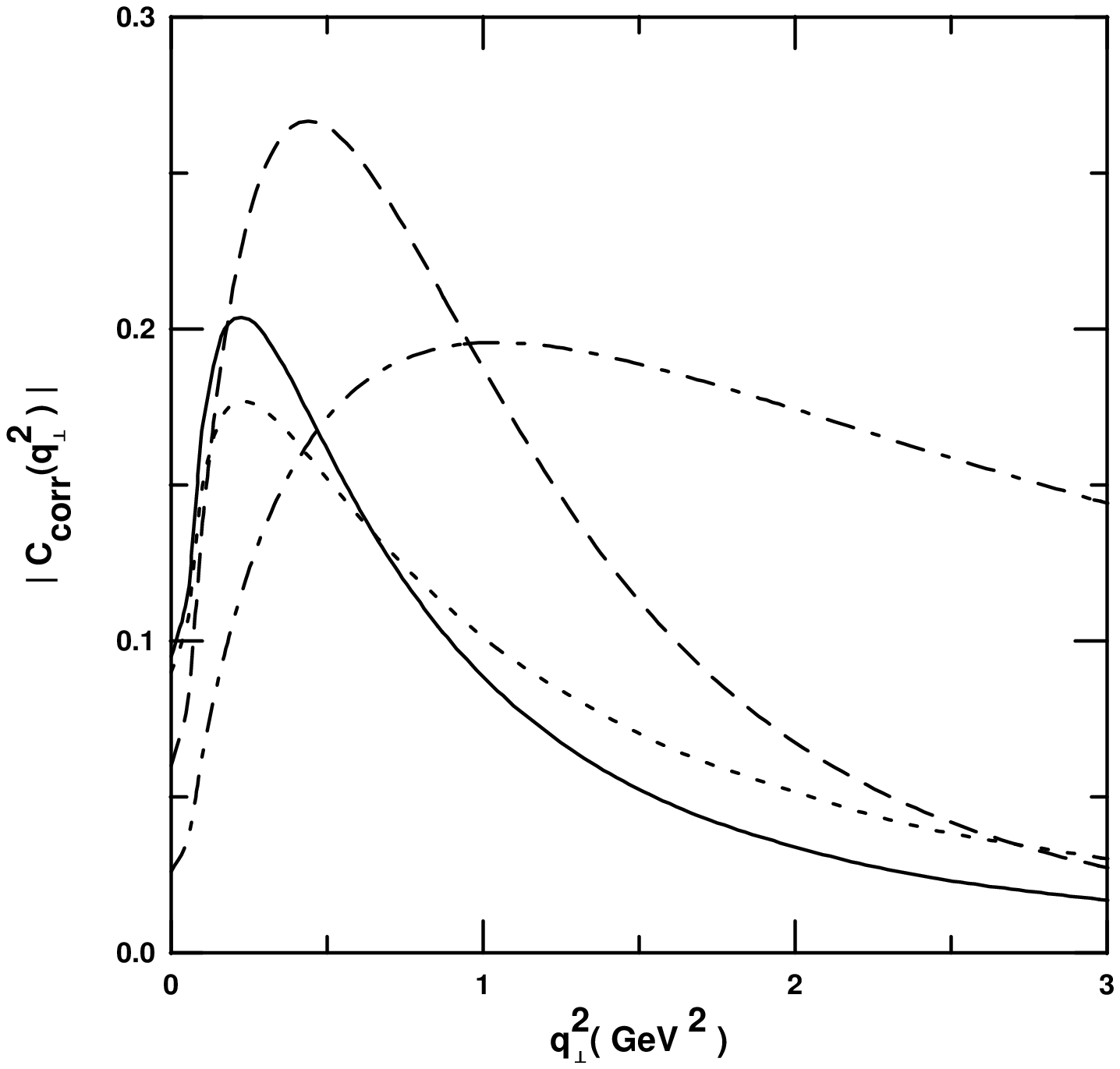}{0.7}
\vskip -3cm
{\bf Fig.4} \ The absolute value of the correlation function of 
the pion, $ C_{corr}(q^2_\perp)$, is plotted against $1/q_\perp$ 
(in Fms), showing the corresponding correlation lengths $l_{corr}$ 
of the four models we consider: 
$\approx$~0.42 fm for the instanton and the hydrogen atom models, 
$\approx$~0.30 fm for the Gaussian model and $\approx$~0.20 fm for 
the model of Godfrey and Isgur.
The parametrization and line conventions 
are the same as in fig.\ref{fig2}.
\vskip 1cm
\label{fig4}
\end{figure}

At this point, we have four models for which we calculate the quark -
antiquark density function $C_{q\bar q}(q^2_\perp )$ in the pion.  In
fig.\ref{fig3},  we plot the results obtained for the instanton, the Gaussian
and the hydrogen-atom models. We also show the model of ref.~\cite{god},
for reference, but this model does not fit $f_\pi$.  The observable
$C_{q\bar q}(q^2_\perp )$ has a zero for these three models, and it
depends strongly on the choice of the wave-function.  In fig.\ref{fig4}, we
show the results for the correlation function $C_{corr}$, given by
Eq.(\ref{Ccorr}), for those four models.  The corresponding correlation
lengths, as seen from the maxima of the curves,  are $l_{corr}$ = 0.42
fm for the instanton and the hydrogen atom models, and $l_{corr}$ =
0.30 fm for the Gaussian model.  The model of ref.~\cite{god} gives
$l_{corr}$ = 0.20 fm.  In this figure we observe the model dependence
of $C_{corr}$.

The Gaussian model, which gives $f_\pi$ and fits reasonably the elastic
form-factor experimental data, does not give the same quark - antiquark
density and the correlation functions as the other two models we have
used (the instanton and the hydrogen atom models). This was shown in
figs.\ref{fig3} and \ref{fig4}.  
This is an indication that the correlation between the
quarks in the pion, as seen by $C_{q\bar q}(q^2_\perp)$ or
$C_{corr}(q^2_\perp)$, can bring more physical information about the
wave-functions not completely contained in the elastic form factor
data.

In conclusion, we have used a light-front sum-rule  defined
from the $W^{++}$ component of structure tensor for inelastic electron
scattering, by integrating it in the $q^-$ component of the momentum
transfer for $q^+=0$. This sum-rule is the light-front generalization
of the non-relativistic Coulomb sum-rule, and it allows to study the
quark-antiquark correlation function in the pion. We made some 
relativistic model calculations and it turned out that this function 
is very sensitive to the light-front model of the pion bound-state 
wave-function.  In general, it can be a useful source of information on the 
relativistic constituent quark wave-function of the hadrons.  
Further investigation about spin - flavor correlations in proton and 
deuteron can provide very important informations about the hadron structure, 
with a better interpretation of existing exclusive and DIS data on spin and 
flavor content of hadrons; it also can suggest an answer to the
intriguing question on the role of nonvalence degrees of freedom.

\section*{Acknowledgements}
The authors are thankful to Prof. S.B. Gerasimov for valuable discussion
on the subject of this work. \ Our thanks for partial financial support 
to Funda\c c\~ao de Amparo \`a Pesquisa do Estado de S\~ao Paulo, 
Conselho Nacional de Desenvolvimento Cient\'{\i}fico e Tecnol\'ogico and 
CAPES of Brasil. \ AD would like also to thank  Russian Academy Science 
grants RFFI-96-02-18096, RFFI-96-02-18097 and San - Petersburg center 
for fundamental research grant: 95-0-6.3-20  for partial support of this work.

\end{document}